**Ontogeny of aerial righting and wing flapping in juvenile birds**
(short title: Aerial righting in juvenile birds)

Dennis Evangelista[*1,2], Sharlene Cam[1], Tony Huynh[1], Igor Krivitskiy[1], and Robert Dudley[1,3]

[1]Department of Integrative Biology, University of California, Berkeley, CA 94720, USA
[2]Department of Biology, University of North Carolina at Chapel Hill, NC 27599, USA
[3]Smithsonian Tropical Research Institute, Balboa, Panama
* author for correspondence (devangel77b@gmail.com)

**Mechanisms of aerial righting in juvenile Chukar Partridge (*Alectoris chukar*) were studied from hatching through 14 days-post-hatching (dph). Asymmetric movements of the wings were used from 1–8 dph to effect progressively more successful righting behaviour via body roll. Following 8 dph, wing motions transitioned to bilaterally symmetric flapping that yielded aerial righting via nose-down pitch, along with substantial increases in vertical force production during descent. Ontogenetically, the use of such wing motions to effect aerial righting precedes both symmetric flapping and a previously documented behaviour in chukar (i.e., wing-assisted incline running) hypothesized to be relevant to incipient flight evolution in birds. These findings highlight the importance of asymmetric wing activation and controlled aerial manoeuvres during bird development, and are potentially relevant to understanding the origins of avian flight.**

**1. Introduction**
In terrestrial vertebrates and invertebrates alike, controlled aerial behaviour can occur even in the absence of obvious aerodynamic surfaces, and includes such phenomena as directed aerial descent [1,2] and the righting reflex [3]. During such manoeuvres, both axial and appendicular structures are used to control body posture via aerodynamic and, in some cases, inertial forces. From the perspective of flight evolution, such controlled aerial behaviours are of particular interest in that they represent potential precursors to the acquisition of fully powered flapping flight using complete wings [4,5]. These transitional forms of aerial manoeuvring can potentially be studied using fossil morphologies within a phylogenetic context [6,7], comparatively among extant volant taxa, or across an ontogenetic sequence for an individual species. In this latter context, the capacity for aerial righting is likely to occur early if not first in the development of flight [5], with controlled and symmetric flapping appearing later as wings and their kinematic activation become more pronounced.

     One well-publicized example of ontogenetic change in avian wing function concerns the use by juvenile chukar and other taxa of symmetric flapping to assist in cursorial ascent of high-angle terrain (wing-assisted incline running, or WAIR [8–12]). This behaviour, which in juvenile chukar occurs developmentally prior to full weight offset via powered flapping, has been proposed as a pathway for wing evolution in the theropod precursors to birds [8,9]. The onset of WAIR in chukar is variously reported to occur at 4 dph (success rate not reported) [8] or at 8–9 dph (three birds) [11], whereas earlier aerial behaviours have not been reported. Although the relevance of modern avian ontogenies to Mesozoic flight origins in birds has been questioned [13], it is nonetheless of interest to assess the aerodynamic capacities of hatchling chukar and early-stage juveniles to place the occurrence of WAIR within a broader functional context. Here, we evaluate the ontogenetic trajectory of aerial righting in juvenile chukar, given the hypothesized evolutionary role of this behaviour relative to subsequent manoeuvring and flight performance [4,5].

## 2. Material and methods

Twenty-six Chukar Partridges (*Alectoris chukar*) were obtained in five separate batches by hatching eggs (Fall Creek Game Birds; Felton, CA) in an incubator. Birds were housed in 53×38×30 cm brooder bins heated with two 100 W flood lamps to maintain air temperatures of 29°C (week 1), 27°C (week 2), and 24°C for subsequent weeks. Birds were kept on wood shavings, and consumed chick starter rations (Purina, St Louis, MO) and water *ad libitum*, as well as grit, mealworms, and fresh grass. Bird mass was recorded daily; the extended wings were also photographed daily for measurement of wing length, area, aspect ratio, and second moment of area (Supplementary Material, table S1). All experiments and husbandry of eggs and chicks followed protocols approved by the Animal Care and Use Committee of the University of California, Berkeley.

We analysed aerial righting behaviours by dropping individuals upside-down in still air from heights between 0.5-1 m within an enclosure shielded from ventilation. The ensuing descent was filmed using high-speed video cameras (either AOS, Baden Daettwil, Switzerland, or Hispec, Fastec Imaging, San Diego, CA) operated at 500 frames/s. Illumination was provided by sixteen 100 W flood lamps hung above the enclosure, air temperatures within which were the same as within the brooder. For filming, a single camera was used with an in-frame scale for calibration oriented parallel to the plane of bird motion. Over the age range tested for aerial righting (1–14 dph), net body motions were primarily vertical with only slight lateral displacements.

Aerial righting trials were scored by two observers simultaneously recording if birds landed on their feet. The landing platform was a loosely draped elastic cloth. Video recordings were used to confirm observer assessments and to diagnose the mode of righting (rolling or pitching; see Results). Bird centre of mass position through time was digitized using the MtrackJ plugin [14] in ImageJ (NIH, Bethesda, MD), and a maximum likelihood estimation routine (`bbmle`) implemented in R [15] was used to quantify the mean aerodynamic force produced during descent (see Supplementary Material).

A total of 26 birds was used in aerial righting trials. For individuals at any given age (i.e., dph), a range of 1–15 drops was obtained from 4–17 birds. Changes in righting success, mode used in righting, and vertical force production among ages were tested using a linear mixed-effects model [15] with age as a fixed factor and individual as a random factor, accounting for repeated measures of individuals; results are plotted for the pooled samples across individuals. The same birds as used in aerial righting tests were also tested at 4, 6, 7, 12, and 19 dph to check for the occurrence of WAIR (at incline angles ranging from 15º to 90º) using methods described elsewhere [8].

## 3. Results

Individuals dropped from an inverted position prior to 9 dph used primarily asymmetric wing and leg movements to roll the body into an upright posture, typically within multiple wing beats (mean ± 1 s.d.: 3.1 ± 1.2 wing beats; figure 1(a)). Typically in such righting behaviour, one wing flapped with large-amplitude movements while the other wing made only small out-of-phase motions; asymmetric scissoring of the legs was also observed. Around 9 dph, the righting mechanism shifted from use of asymmetric wing and leg motions to symmetric flapping of the two wings to effect nose-down pitching (figure 1(b)). This behaviour was typically completed within a single wing beat (mean ± 1 s.d.: 1.0 ± 0.3 wing beats). Both wings flapped in phase at high amplitude; leg movements were either absent or confined to a single in-phase kick of both legs.

Righting success rates improved quickly in juvenile chukar ($t=15.4$, d.f.=112, $p<1\times10^{-6}$), reaching 50% at 4 dph and 100% after 9 dph (figure 2(a)), in tandem with the shift to righting via rolling to changes in pitch (figure 2(b); $t=-2.23$, d.f.=57, $p=0.0296$). In parallel, vertical aerodynamic forces (normalized to body weight) increased with the onset of symmetric wing flapping (figure 2(c); $t=9.06$, d.f.=36, $p<1\times10^{-6}$).

At 4 dph, WAIR could not be elicited from any bird ($N=8$) at any tested incline angle. We observed a single weak bout of WAIR in one of eight birds tested at 6 and at 7 dph at an incline angle below 45º. At 12 dph, one of four birds exhibited WAIR at 45º, whereas five of eight birds exhibited this behaviour at 19 dph at 45º. Overall, birds were reluctant to engage in WAIR at any angle, and tended instead to run up the incline obliquely using their legs alone, or to jump and otherwise avoid the incline.

**4. Discussion**
Chukar grew progressively better at responding to the aerial challenge of righting from an initially upside-down posture. At 1 dph, birds did little to alter their fall as compared to a passive falling projectile, but even at this early stage could occasionally right themselves (25%, figure 2(a)). However, aerial righting developed rapidly (50% at 4 dph, figure 2(a)) via asymmetric flapping and induced rolling, and was fully developed by 9 dph with the onset of symmetric flapping. The former behaviour is concurrent with an asymmetric quadrupedal crawling stage that occurs before WAIR [11]. The latter, and associated vertical forces (figure 2(c)), is concurrent with the flapping descent described previously for juvenile chukar [11].

The capacity for aerial righting as described here well preceded the onset of WAIR. Prior studies of chukar report different timings for the onset of WAIR [8,11]. Here, WAIR up a 45º incline was first observed in one bird at 12 dph; WAIR was only observed in a majority of birds at 19 dph. By 12 dph, birds had already transitioned to symmetric wing flapping and righting via associated changes in body pitch (figure 1(b)). The regular occurrence of WAIR under natural circumstances has yet to be documented in juvenile chukar or other birds, and we accordingly encourage further study of this fascinating behaviour in realistic ecological settings.

Aerial righting, even with asymmetric wing motions, requires precise control of rotational moments to effect a desired body posture, and as such should be viewed as a sophisticated manoeuvre. Its occurrence early in chukar ontogeny, and well prior to the ontogenetic advent of WAIR, suggests a broader diversity of controlled aerial behaviours in juvenile birds than has been previously recognized. This finding, in and of itself, is of no necessary evolutionary significance relative to avian flight origins [13]. Prior studies of WAIR have nonetheless concluded that aerodynamic force production via a "fundamental" symmetric wing stroke aligned to gravity must have preceded the capacity for aerial manoeuvring and flight control [8,9]. These studies similarly ascribe to basal taxa only limited flight styles accomplished solely by adjusting power output, with bilaterally asymmetric modifications to the wing stroke for "advanced" forms of aerial locomotion only ocurring in derived taxa [8]. In contrast to this perspective, the capacity for aerial righting via asymmetric wing motions clearly occurs very early in chukar development, and is but one of potentially many uses of incipient wings in birds. Documentation of additional maneuvering abilities early in development, such as pitching, yawing, or aerial displacement towards targets of interest would add further indirect support to this hypothesis. Palaeontological documentation of hind wings in ancestral paravian taxa [16] further suggests a concurrent diversity of aerodynamic function [6]. More generally, manoeuvring is clearly important for any animal exhibiting rudimentary aerial capacity [7], even in the complete absence of wings [1–5]. Righting by chukar at only 1 dph exemplifies this important functional demand.

**Acknowledgements**
We thank L Waldrop, K Dorgan, Y Munk, Y Zeng, E Kim, M Badger, S Chang, N Sapir, V Ortega, M Wolf, J McGuire, and R Fearing for assistance and comments. We also thank M Ho, Y Lin, R Stevenson, J Ye and the Berkeley Undergraduate Research Apprentice Program; L Guillen and K Moorhouse for advice in raising birds; and the Berkeley Centre for Integrative Biomechanics Education and Research for the use of high speed cameras.

**Data accessibility**
Data [17] uploaded to Dryad: http://doi.org/10.5071/dryad.s6k44

**Funding statement**
DE was supported by NSF Integrative Graduate Education and Research Traineeship (IGERT) #DGE-0903711 and by grants from the Berkeley Sigma Xi chapter and the national Sigma Xi.

Figure 1. Representative righting manoeuvres in Chukar Partridge (images separated by 100 ms). (a) Righting via roll using asymmetric wing and leg movements, used prior to 14 dph (bird #25, 9 dph); (b) righting via pitch using symmetric wing movements, prevalent after 9 dph (bird #42, 10 dph).

Figure 2. (a) Percent righting ($N$=26 birds, number of drops as indicated) and (b) righting mode ($N$=26 birds, number of successful rightings as indicated), and (c) vertical force production ($N$=5 birds, except for $N$=1 at 14 dph; data represent mean ± 1 s.d.) versus age in Chukar Partridge. Righting via roll, as accomplished by asymmetric wing and leg movements, is used prior to 14 dph. Around 9 dph, birds switch to righting via pitch using symmetric wing motions, and vertical force production increases concomitantly. See text for statistical results.

a

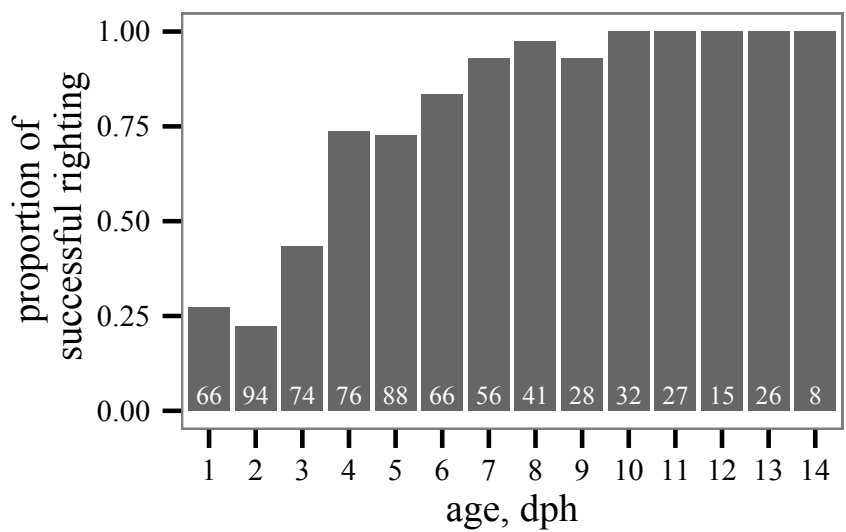

b

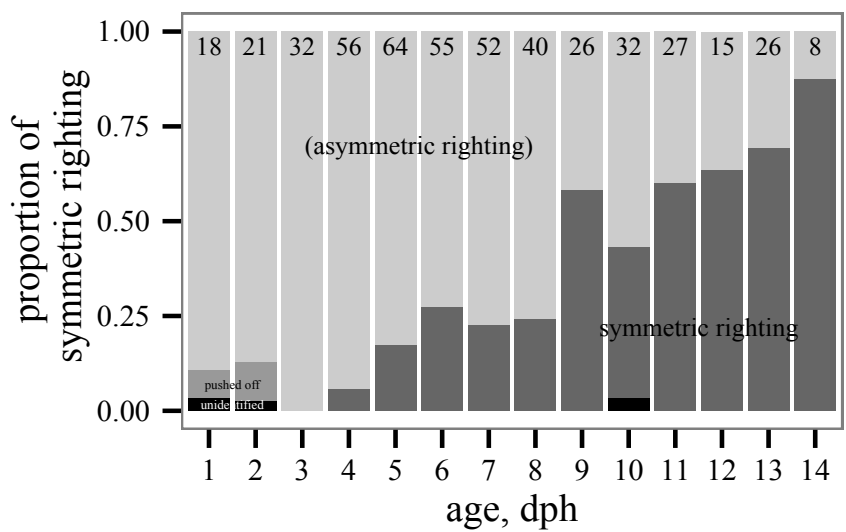

c

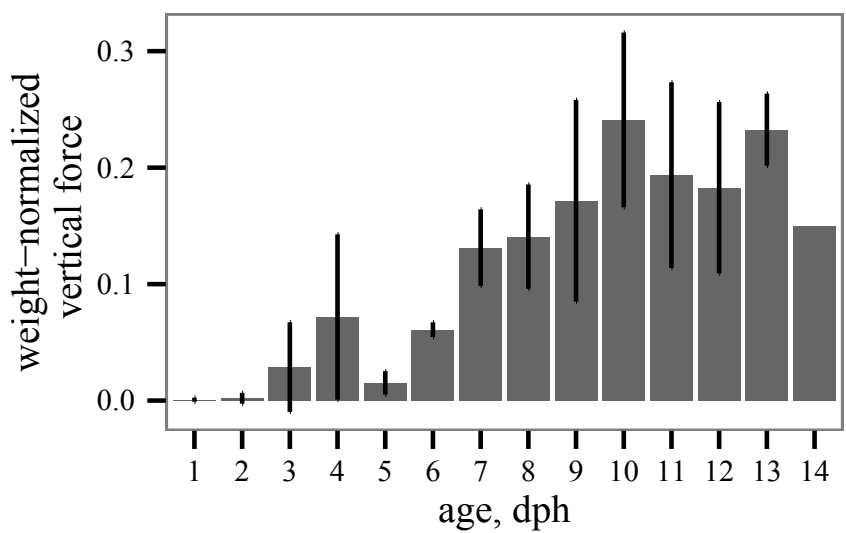